# LSTM Recurrent Neural Networks for Cybersecurity Named Entity Recognition

Houssem Gasmi[1,2], Abdelaziz Bouras[1]
[1]Computer Science Department, College of Engineering, Qatar University
Doha, Qatar
[2]Université Lumière Lyon 2, Lyon, France
email:houssem.gasmi@qu.edu.qa
email:abdelaziz.bouras@qu.edu.qa

Jannik Laval
DISP Laboratory
Université Lumière Lyon 2
Lyon, France
email:jannik.laval@univ-lyon2.fr

*Abstract*— The automated and timely conversion of cybersecurity information from unstructured online sources, such as blogs and articles to more formal representations has become a necessity for many applications in the domain nowadays. Named Entity Recognition (NER) is one of the early phases towards this goal. It involves the detection of the relevant domain entities, such as product, version, attack name, etc. in technical documents. Although generally considered a simple task in the information extraction field, it is quite challenging in some domains like cybersecurity because of the complex structure of its entities. The state of the art methods require time-consuming and labor intensive feature engineering that describes the properties of the entities, their context, domain knowledge, and linguistic characteristics. The model demonstrated in this paper is domain independent and does not rely on any features specific to the entities in the cybersecurity domain, hence does not require expert knowledge to perform feature engineering. The method used relies on a type of recurrent neural networks called Long Short-Term Memory (LSTM) and the Conditional Random Fields (CRFs) method. The results we obtained showed that this method outperforms the state of the art methods given an annotated corpus of a decent size.

I. INTRODUCTION

Timely extraction of cybersecurity information from diverse online web sources, such as news, vendor bulletins, blogs, forums, and online databases is vital for many types of applications. One important application is the conversion of unstructured cybersecurity information to a more structured form like ontologies. Knowledge modeling of cyber-attacks for instance simplifies the work of auditors and analysts [1]. At the heart of the information extraction tasks is the recognition of named entities of the domain, such as vendors, products, versions, or programming languages. The current NER tools that give the best performance in the field are based on feature engineering. These tools rely on the specific characterizing features of the entities in the field, for example, a decimal number that follows a product is very likely to be the version of that product and not quantities of it. A sequence of words starting with capital letters is likely to be a product name rather than a company name and so on.

Feature engineering has many issues and limitations. Firstly, it relies heavily on the experience of the person and the lengthy trial and error process that accompanies that. Secondly, feature engineering relies on look-ups or dictionaries to identify known entities [2]. These dictionaries are hard to build and harder to maintain especially with highly dynamic fields, such as cybersecurity. These activities constitute the majority of the time needed to construct these NER tools. The results could be satisfactory despite requiring considerable maintenance efforts to keep them up to date as more products are released and written about online. However, these tools are domain specific and do not achieve good accuracy when applied to other domains. For instance, a tool that is designed to recognize entities in the biochemistry field will perform very poorly in the domain of cybersecurity [3].

CRFs emerged in recent years as the most successful and de facto standard method for entity extraction. In this paper, we show that a domain agnostic method that is based on the recent advances in the deep learning field and word embeddings outperforms traditional methods, such as the CRFs. The first advancement, which is the word2vec word embedding method was introduced by Mikolov et al. [4] . It represents each word in the corpora by a low dimensional vector. Besides the gain in space, one of the main advantages of this representation compared to the traditional one-hot vectors [5] is the ability of these vectors to reflect the semantic relationship between the words. For instance, the difference between the vectors representing the words 'king' and 'queen' is similar to the difference between the vectors representing the words 'man' and 'woman'. These relationships result in the clustering of semantically similar words in the vector space. For instance, the words 'IBM' and 'Microsoft' will be in the same cluster, while words of products like 'Ubuntu' and 'Web Sphere Server' appear together in a different cluster.

The second advancement is the recent breakthroughs in the deep learning field. It became feasible and practical because of the increase in the hardware processing power especially GPUs and the surge in the data available for training. Deep neural networks can automatically learn non-linear combinations of features with enough training data. Hence, they alleviate the user from the time-consuming feature engineering [6]. Besides requiring feature engineering, traditional methods such as CRFs can only learn linear combinations of the defined features. The specific deep learning method we used is LSTM, which is a type of Recurrent Neural Networks (RNNs) that are particularly suitable for processing sequences of data, such as time series and natural language text [7].

We applied the LSTM-CRF architecture suggested by Lampal et al. [8] to the domain of cybersecurity NER. This architecture combines LSTM, word2vec models, and CRFs. The main characteristic of this method is that it is domain and entity type agnostic and can be applied to any domain. All it needs as input is an annotated corpus in the same format as the CoNLL-2000 dataset [9]. We compared the performance of LSTM-CRF with one of the fastest and most accurate CRF implementations, which is CRFSuite [10]. Unlike domains such as the biomedical domain, annotated corpora in the field of cybersecurity are not widely available. The corpora used to train the model were generated as part of the work of Bridges et al [1]. LSTM-CRF achieved 2% better overall item accuracy than the CRF tool.

The paper is organized as follows: Section II reviews the related work in the field. Section III provides an overview of the LSTM-CRF model. The next Section describes our evaluation method and the data pre-processing steps. Section V outlines and discusses the results. Finally, Section VI concludes the paper.

## II. RELATED WORK

Approaches to NER are mainly either rule-based or machine learning/statistical-based [11], although quite often the two techniques are mixed [12]. Rule-based methods typically are a combination of Gazette-based lookups and pattern matching rules that are hand-coded by a domain expert. These rules use the contextual information of the entity to determine whether candidate entities from the Gazette are valid or not. Statistical based NER approaches use a variety of models, such as Maximum Entropy Models [13], Hidden Markov Models (HMMs) [14], Support Vector Machines (SVMs) [15], Perceptrons [16], Conditional Random Fields (CRFs) [17], or neural networks [18]. The most successful NER approaches include those based on CRFs. CRFs address the NER problem using a sequence-labeling model. In this model, the label of an entity is modeled as dependent on the labels of the preceding and following entities in a specified window. Examples of frameworks that are available for CRF-based NER are Stanford NER and CRFSuite.

More recently, deep neural networks have been considered as a potential alternative to the traditional statistical methods as they address many of their shortcomings [19]. One of the main obstacles that prevent the adoption of the methods mentioned above is feature engineering. Neural networks essentially allow the features to be learned automatically. In practice, this can significantly decrease the amount of human effort required in various applications. More importantly, empirical results across a broad set of domains have shown that the learned features in neural networks can give very significant improvements in accuracy over hand-engineered features. RNNs, a class of neural networks have been studied and proved that they can process input with variable lengths as they have a long time memory. This property resulted in notable successes with several NLP tasks like speech recognition and machine translation [20]. LSTM further improved the performance of RNNs and allowed the learning between arbitrary long-distance dependencies [21].

Various methods have been applied to extract entities and their relations in the cybersecurity domain. Jones et al. [22] implemented a bootstrapping algorithm that requires little input data to extract security entities and the relationship between them from the text. An SVM classifier has been used by Mulwad et al. [23] to separate cybersecurity vulnerability descriptions from non-relevant ones. The classifier uses Wikitology and a computer security taxonomy to identify and classify domain entities. The two previously mentioned works relied on standard NER tools to recognize the domain concepts. While these NER tools obtained satisfactory results in general texts, such as news, they performed poorly when applied to more technical domains, such as cybersecurity because these tools are not trained on domain-specific concept identification. For instance, the Stanford NER tool is trained using a training corpus consisting mainly of news documents that are largely annotated with general entity types, such as names of people, locations, organizations, etc.

To overcome the limitations of NER tools in technical domains and identify mentions of domain-specific entities, Goldberg [5] adopted an approach that trains the CRF classifier of the Stanford NER framework on a hand-labeled training data. He achieved acceptable results that are much better than the two previous efforts. Although they produced good results, the effort involved in painstakingly annotating even a small corpus prohibits the practical implementation of this approach. To address this problem, Joshi et al. [3] developed a method to automate the labeling of training data when there is no domain-specific training data available. The labeling process leverages several data sources by combining several related domain-specific structured data to infer entities in the text. Next, a Maximum Entropy Markov Model has been trained on a corpus of nearly 750,000 words and achieved precisions above 90%. This type of training relies on external sources for corpus annotation. These resources need to be regularly maintained and updated to maintain the quality and precision of the text labeling.

Given the benefits of neural networks, this paper aims to apply the LSTM method on the problem of NER in the cybersecurity domain using the corpora made available by Joshi et al. [3]. We analyzed the results achieved and compared them with the CRF method.

### III. LSTM-CRF MODEL

In this Section, we will provide an overview of the LSTM-CRF architecture as presented by Lample et al. [8].

#### A. LSTM-CRF Model

RNNs are neural networks that have the capability to detect and learn patterns in data sequences. These sequences could be natural language text, spoken words, genomes, stock market time series, etc. Recurrent networks combine the current input (e.g., current word) with the previous perceived input (earlier words in the text). However, RNNs are not good at handling long-term dependencies. When the previous input becomes large, RNNs suffer from the vanishing or exploding gradient problems. They can also be challenging to training and very unlikely to converge when the number of parameters becomes large.

LSTMs were first introduced by Hochreiter et al. [7] They are an improvement on RNNs and can learn arbitrary long-term dependencies, hence can be used for a variety of applications such as natural language processing and stock market analysis. LSTMs have a similar chain structure as RNNs, but the structure of the repeating nodes is different. LSTMs have multiple layers that communicate with each other in a particular way. A typical LSTM consists of an input gate, an output gate, a memory cell, and a forget gate. Briefly, these gates control which input to pass to the memory cell to remember it in the future and which earlier state to forget. The implementation used is as follows [8]:

$$\mathbf{i_t} = \sigma(\mathbf{W}_{xi}\mathbf{x}_t + \mathbf{W}_{hi}\mathbf{h}_{t-1} + \mathbf{W}_{ci}\mathbf{c}_{t-1} + \mathbf{b}_i)$$
$$\mathbf{c}_t = (1 - \mathbf{i}_t) \odot \mathbf{c}_{t-1} + \mathbf{i}_t \tanh(\mathbf{W}_{xc}\mathbf{x}_t + \mathbf{W}_{hc}\mathbf{h}_{t-1} + \mathbf{b}_c)$$
$$\mathbf{o}_t = \sigma(\mathbf{W}_{xo}\mathbf{x}_t + \mathbf{W}_{ho}\mathbf{h}_{t-1} + \mathbf{W}_{co}\mathbf{c}_t + \mathbf{b}_o)$$
$$\mathbf{h}_t = \mathbf{o}_t \odot \tanh(\mathbf{c}_t)$$

The sigma sign σ is the elementwise sigmoid function and ⊙ is the elementwise product.

Assuming we have a sequence of *n* words X = ($x_1$, $x_2$,…, $x_n$) and each word is represented by a vector of dimension d. LSTM computes the left context $lh_t$ which represents all the words that precede the word *t*. A right context $rh_t$ is also computed using another LSTM that reads the same text sequence in reverse order by starting from the end and go backward. This technique proved very useful and the resulting architecture, which consists of a forward LSTM and a backward LSTM, is called a Bidirectional LSTM. The resulting representation of a word is obtained by concatenating the left and right contexts to get the representation $h_t$ = [$lh_t$;$rh_t$]. This representation is useful for various tagging applications, such as the NER problem at hand in this paper.

Figure 1 shows the architecture of the Bidirectional LSTM-CRF model. It consists of three layers.

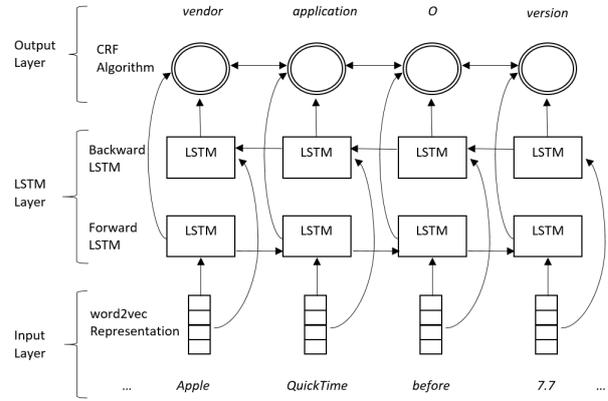

Figure 1. Bidirectional LSTM Architecture

From the bottom, the first layer is the embedding layer. This layer takes as input the sequence S of words $w_1, w_2, ..., w_t$ and emits a dense vector representation (embedding) $x_t$ for each of the words in the sequence. The sequence of embeddings $x_1, x_2, ... , x_t$ is then passed to the bi-directional LSTM layer which refines the input and feeds it to the final CRF layer. In the last layer, the Viterbi algorithm is applied to generate the output of the neural network, which represents the most probable tag for the word.

### IV. EVALUATION

In this section, we will introduce the benchmark tool, the preprocessing performed on the gold standard corpora, and the metrics we used for evaluation.

#### A. Competitor System

We compare the performance of the LSTM-CRF architecture against a CRF tool that uses a generic feature set for NER with word embeddings. These features were designed for domain-independent NER and defined by the tool writer. Using word embeddings in both systems will help us compare only the CRF method with the suggested LSTM-CRF architecture and negate the effect of word embeddings. We used the CRFSuite to train a CRF model using the default settings of the tool.

#### B. Gold standard corpora

We performed our evaluation on around 40 entity types defined in three corpora and also analyzed the performance of the model on a subset of the seven most significant entities of the domain. Each word in these corpora is auto-annotated with an entity type. The corpus is an auto-labeled cyber security domain text that was generated for use in the Stucco project. It includes all descriptions from CVE/NVD entries starting in 2010, in addition to entries from MS Bulletins and Metasploit. As stated in [1]: "While labelling these descriptions may be useful in itself, the intended purpose of this corpus is to serve as training data for a supervised learning algorithm that accurately labels other text documents in this domain, such as blogs, news articles, and tweets.".

*C. Text Preprocessing*

In its original form as provided by Bridges et al [1], all the corpora were stored in a single JSON file with each corpus represented by a high-level JSON element. To facilitate further processing, we converted the file to the CoNLL2000 format as the input for the LSTM-CRF model. In the newly single annotated corpus, we removed the separation between each of the three corpora and annotated every word in a separate line. Each line contains the word mentioned in the text and its entity type as show in the following example:

...
*Apple B-vendor*
*QuickTime B-application*
*before B-version*
*7.7 I-version*
*allows B-relevant_term*
*remote B-relevant_term*
*attackers I-relevant_term*
*to O*
...

As for the CRF model, the CRFSuite requires the training data to be in the CoNLL2003 format that includes the Part of Speech (POS) and chunking information with the NER tag appearing first as shown below:

...
*B-vendor Apple NNP O*
*B-application QuickTime NNP O*
*B-version before IN O*
*I-version 7.7 CD O*
*B-relevant_term allows NNS O*
*B-relevant_term remote VBP O*
*I-relevant_term attackers NNS O*
*O to TO O*
...

As the original corpus did not contain the POS and chunking information, the training corpus had to be reprocessed. We started by converting it to its original form (i.e., a set of paragraphs). Then, we used the python NLTK library to extract the necessary information for each word in the corpus. Finally, we converted the text back to the expected format shown above.

*D. Evaluation Metrics*

We divided the annotated corpus into 3 disjoint subsets. 70% was allocated for the training of the model, 10% for the holdout cross-validation set (or development), and 20% for the evaluation of the model. We compared the two models (LSTM-CRF and CRF) in terms of accuracy, precision, recall, and F1-score for the full set of tags and for a subset of the most relevant tags of the domain. In our experiments, the hyperparameters of the LSTM-CRF model were set to the default values used by Lample et al. [8].

V. RESULTS AND DISCUSSION

We evaluated the performance of the NER method that is based on the LSTM-CRF architecture against a traditional state of the art CRF tool that uses standard NER features. The evaluation was performed on three different sets covering over 40 entity types from the cybersecurity domain. For evaluation purposes, we will analyze the average performance of models across all the entity types, then we will consider the most popular entities that appear frequently in the cybersecurity vulnerability descriptions and evaluate the performance on these entities only. The entities considered are *vendor, application, version, file, operating system (os), hardware,* and *edition*. The reason for this is that we are usually not interested in extracting all entity types but only a subset of them that are most relevant to the application at hand.

*A. Performance of LSTM-CRF and CRF*

Starting with the global item accuracy of both models, Figure 2 shows the accuracy values measured on the test set at each iteration of the training stage for 100 iterations. LSTM-CRF achieved an accuracy of 95.8% after the first iteration and increased gradually to reach values between 98.2% and 98.3% starting from iteration 23 until the end of the training. On the other hand, the CRF method started slowly at accuracies of 65% and increased rapidly to reach accuracies of 96% where it leveled off to reach eventually 96.35% at the end of the training.

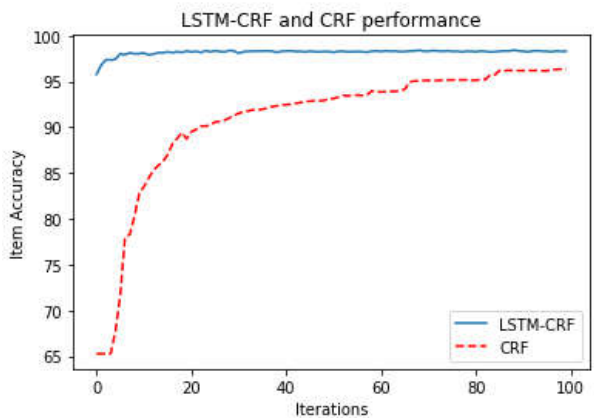

Figure 2. Item Accuracy for LSTM-CRF and CFR

The average performance of the two models across all the entity types in the training set is shown in the table below:

TABLE I. AVERAGE PERFORMANCE METRICS FOR ALL ENTITY TYPES

|  | Precision (%) | Recall (%) | F1-score (%) |
|---|---|---|---|
| LSTM-CRF | 85.16 | 80.70 | 83.37 |
| CRF | 80.26 | 73.55 | 75.97 |

As we can see, the performance metrics in terms of precision, recall, and F1-score show that the results for LSTM-CRF are better that their CRF counterparts.

ICSEA 2018 : The Thirteenth International Conference on Software Engineering AdvancesWe then compared the performance of LSTM-CRF and CRF on the most popular seven entity types in the domain. The results for each method for the different entities in terms of F1-score, precision, and recall are shown in Tables II, III, and IV. LSTM-CRF achieved the best performance for all the entities types with the exception of the *hardware*, and *edition* tags, which affected the average considerably. On average (macro average), F1-scores are 82.8% for the generic LSTM-CRF method and 84.4% for the generic CRF method. In terms of precision, the results are close at 87.2% and 89% respectively. As for the recall, it is 80.1% and 81.8% respectively.

TABLE II. F1-SCORES OF CRF AND LSTM-CRF FOR SEVEN ENTITY TAGS

| Entity type | LSTM-CRF (%) | | CRF (%) |
|---|---|---|---|
| | *Development* | *Test* | |
| *Vendor* | 94 | 93 | 92 |
| *Application* | 90 | 89 | 87 |
| *Version* | 98 | 98 | 95 |
| *Edition* | 80 | 60 | 80 |
| *OS* | 95 | 95 | 93 |
| *Hardware* | 60 | 46 | 63 |
| *File* | 99 | 99 | 84 |
| **Average** | **88** | **82.8** | **84.4** |

TABLE III. PRECISION OF CRF AND LSTM-CRF FOR SEVEN ENTITY TAGS.

| Entity type | LSTM-CRF (%) | | CRF (%) |
|---|---|---|---|
| | *Development* | *Test* | |
| *Vendor* | 95 | 94 | 94 |
| *Application* | 90 | 89 | 88 |
| *Version* | 98 | 98 | 95 |
| *Edition* | 80 | 76 | 87 |
| *OS* | 98 | 97 | 95 |
| *Hardware* | 69 | 57 | 79 |
| *File* | 99 | 1 | 85 |
| **Average** | **89.8** | **87.2** | **89** |

TABLE IV. RECALL OF CRF AND LSTM-CRF FOR SEVEN ENTITY TAGS

| Entity type | LSTM-CRF (%) | | CRF (%) |
|---|---|---|---|
| | *Development* | *Test* | |
| *Vendor* | 93 | 92 | 90 |
| *Application* | 89 | 90 | 86 |
| *Version* | 98 | 98 | 95 |
| *Edition* | 79 | 50 | 75 |
| *OS* | 95 | 93 | 91 |
| *Hardware* | 54 | 39 | 52 |
| *File* | 100 | 99 | 84 |
| **Average** | **86.8** | **80.1** | **81.8** |

We can see that the overall item accuracy of the resulting LSTM-CRF model is higher by 2% than the CRF model. Likewise, the average precision, recall, and F1-score across all entity types are better by an average of 6.5%. As for the performance metrics per entity type, the LSTM-CRF model performed better on five entity types and poorly on the *hardware* and *edition* tags. The reason for this poor performance is related to the size of the training data. Deep learning algorithms such as LSTM, needs lots of data for better predictions. The more data we have, the better the prediction model can get. Upon analyzing the data set, it turned out that very few entities are tagged with these two tags compared to the other entities. There are 549 entities tagged as *hardware* and 565 tagged as *edition*. These numbers are relatively low compared to other tags, such as *application* (19093 tags) and *vendor* (10518 tags). Therefore, the first five tags overwhelmed the other poorly performing two tags. Increasing the size of the training data that contains more examples of these tags will improve the prediction of the model.

VI. CONCLUSION AND FUTURE WORK

As this paper showed, the results demonstrate that LSTM-CRF improved the accuracy of NER extraction over the state-of-art traditional pure statistical CRF method. What is impressive about the LSTM-CRF method is that it does not require any feature engineering and is entirely entity type agnostic. Even the format of the training corpus is much simpler, thus requiring less text pre-processing. This alleviates the need to develop domain-specific tools and dictionaries for NER. In the future, our research will concentrate on applying the LSTM-CRF method on entity Relations Extraction (RE). RE is concerned with attempting to find occurrences of relations among domain entities in text. This would provide a better understanding of product vulnerability descriptions. For example, RE could extract information from a vulnerability description that would help us distinguish between the product or tool that is the mean of an attack and the product being attacked. With information extraction becoming more accurate, more automated, and easier to achieve using recent neural networks advancements, there is a pressing need to turn this advancement into applications in the domain of cybersecurity. One such application is the conversion of the textual descriptions of cybersecurity vulnerabilities that are available in the web into a more formal representation like ontologies. This gives cybersecurity professionals the necessary tools that grant them rapid access to the information needed for decision-making.

ACKNOWLEDGEMENTS
ACKNOWLEDGEMENTS

This publication was made possible by NPRP grant # NPRP 7-1883-5-289 from the Qatar National Research Fund (a member of Qatar Foundation). The statements made herein are solely the responsibility of the authors.

REFERENCES

[1] R. A. Bridges, C. L. Jones, M. D. Iannacone, K. M. Testa, and J. R. Goodall, "Automatic labeling for entity extraction in cyber security," *arXiv preprint arXiv:1308.4941*, 2013.

[2] T. H. Nguyen and R. Grishman, "Event detection and domain adaptation with convolutional neural networks," in *Proceedings of the 53rd Annual Meeting of the Association for Computational Linguistics and the 7th International Joint*

*Conference on Natural Language Processing (Volume 2: Short Papers)*, pp. 365–371, 2015.

[3] A. Joshi, R. Lal, T. Finin, and A. Joshi, "Extracting cybersecurity related linked data from text," in *Semantic Computing (ICSC), 2013 IEEE Seventh International Conference on*, pp. 252-259, 2013.

[4] T. Mikolov, I. Sutskever, K. Chen, G. S. Corrado, and J. Dean, "Distributed representations of words and phrases and their compositionality," in *Advances in neural information processing systems*, pp. 3111-3119, 2013.

[5] Y. Goldberg, "A primer on neural network models for natural language processing," *Journal of Artificial Intelligence Research*, vol. 57, pp. 345-420, 2016.

[6] J. Schmidhuber, "Deep learning in neural networks: An overview," *Neural networks*, vol. 61, pp. 85-117, 2015.

[7] S. Hochreiter and J. Schmidhuber, "Long short-term memory," *Neural computation*, vol. 9, pp. 1735-1780, 1997.

[8] G. Lample, M. Ballesteros, S. Subramanian, K. Kawakami, and C. Dyer, "Neural architectures for named entity recognition," *arXiv preprint arXiv:1603.01360*, 2016.

[9] E. F. Tjong Kim Sang and S. Buchholz, "Introduction to the CoNLL-2000 shared task: Chunking," in *Proceedings of the 2nd workshop on Learning language in logic and the 4th conference on Computational natural language learning-Volume 7*, pp. 127-132, 2000.

[10] N. Okazaki, *CRFsuite: a fast implementation of Conditional Random Fields (CRFs)*, 2007.

[11] P. Cimiano, S. Handschuh, and S. Staab, "Towards the self-annotating web," in *Proceedings of the 13th international conference on World Wide Web*, pp. 462-471, 2004.

[12] P. Pantel and M. Pennacchiotti, "Automatically Harvesting and Ontologizing Semantic Relations," in Proceedings of the 2008 Conference on Ontology Learning and Population: Bridging the Gap Between Text and Knowledge, Amsterdam, The Netherlands, The Netherlands, pp. 171-195, 2008.

[13] H. L. Chieu and H. T. Ng, "Named entity recognition: a maximum entropy approach using global information," in *Proceedings of the 19th international conference on Computational linguistics-Volume 1*, pp. 190-196, 2002.

[14] A. McCallum, D. Freitag, and F. C. N. Pereira, "Maximum Entropy Markov Models for Information Extraction and Segmentation.," in *Icml*, pp. 591-598, 2000.

[15] H. Isozaki and H. Kazawa, "Efficient support vector classifiers for named entity recognition," in *Proceedings of the 19th international conference on Computational linguistics-Volume 1*, pp. 1-7, 2002.

[16] X. Carreras, L. Màrquez, and L. Padró, "Learning a perceptron-based named entity chunker via online recognition feedback" in *Proceedings of the seventh conference on Natural language learning at HLT-NAACL 2003-Volume 4*, pp. 156-159, 2003.

[17] A. McCallum and W. Li, "Early results for named entity recognition with conditional random fields, feature induction and web-enhanced lexicons," in *Proceedings of the seventh conference on Natural language learning at HLT-NAACL 2003-Volume 4*, pp. 188-191, 2003.

[18] R. Collobert, J. Weston, L. Bottou, M. Karlen, K. Kavukcuoglu, and P. Kuksa, "Natural language processing (almost) from scratch," *Journal of Machine Learning Research*, vol. 12, pp. 2493-2537, 2011.

[19] Y. Goldberg, "A Primer on Neural Network Models for Natural Language Processing.," *J. Artif. Intell. Res.(JAIR)*, vol. 57, pp. 345-420, 2016.

[20] A. Graves, A. Mohamed, and G. Hinton, "Speech recognition with deep recurrent neural networks," in *Acoustics, speech and signal processing (icassp), ieee international conference on*, pp. 6645-6649, 2013.

[21] F. A. Gers, J. A. Schmidhuber and F. A. Cummins, "Learning to Forget: Continual Prediction with LSTM,", *Neural Compution*, vol. 12, pp. 2451-2471, 10 2000.

[22] L. Jones, R. A. Bridges, K. M. T. Huffer and J. R. Goodall, "Towards a relation extraction framework for cyber-security concepts," in *Proceedings of the 10th Annual Cyber and Information Security Research Conference*, pp. 11, 2015

[23] V. Mulwad, W. Li, A. Joshi, T. Finin and K. Viswanathan, "Extracting information about security vulnerabilities from web text," in *Web Intelligence and Intelligent Agent Technology (WI-IAT), 2011 IEEE/WIC/ACM International Conference on*, 2011.